# Discrete-time Control of Bilateral Teleoperation Systems: A Review

Amir. Aminzadeh. Ghavifekr*, Amir. Rikhtehgar. Ghiasi **, Mohammad Ali. Badamchizadeh ***
* Department of Electrical and Computer Engineering, University of Tabriz, Tabriz, Iran
Email: aa.ghavifekr@tabrizu.ac.ir
** Department of Electrical and Computer Engineering, University of Tabriz, Tabriz, Iran
Email: agiasi@tabrizu.ac.ir
***Department of Electrical and Computer Engineering, University of Tabriz, Tabriz, Iran
Email: mbadamchi@tabrizu.ac.ir

***Abstract:*** The possibility of operating in remote environments using teleoperation systems has been considered widely in the control literature. This paper presents a review on the discrete-time teleoperation systems, including issues such as stability, passivity and time delays. Using discrete-time methods for a master-slave teleoperation system can simplify control implementation. Varieties of control schemes have been proposed for these systems and major concerns such as passivity, stability and transparency have been studied. Recently, unreliable communication networks affected by packet loss and variable transmission delays have been received much attention. Thus, it is worth considering discrete-time theories for bilateral teleoperation architectures, which are formulated on the same lines as the continuous-time systems. Despite the extensive amount of researches concerning continuous-time teleoperation systems, only a few papers have been published on the analysis and controller design for discrete bilateral forms. This paper takes into account the challenges for the discrete structure of bilateral teleoperation systems and notifies the recent contributions in this area. The effect of sampling time on the stability-transparency trade-off and the task performance is taken into consideration in this review. These studies can help to design guidelines to have better transparency and stable teleoperation systems.

## 1. Introduction

During the last decades, various teleoperation systems have been developed to allow the human operators to perform tasks in remote or hazardous environments.
Teleoperation systems have been used for a number of various tasks, e.g., handling harmful or toxic materials, operating in remote areas such as telesurgery and space explorations and performing tasks that require extreme precision like in nuclear reactors. Hokayem and Spong [1] presented a comprehensive overview of researches and developments for the teleoperation systems.
A typical teleoperation system is commonly composed of five interconnected elements: master and slave robots, environment, a human operator and a communication channel. The block diagram of this system is illustrated in Fig. 1.

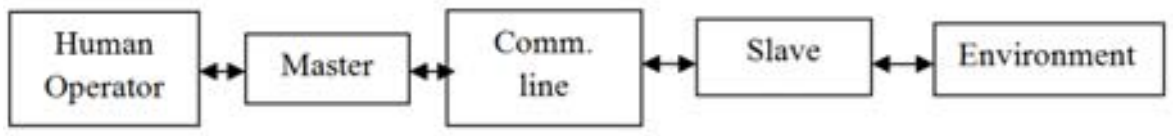

Fig. 1. The schematic of the typical teleoperation system

The operator controls the slave robot to implement given tasks in a remote manner. The slave should follow the movement of the master robot. In a bilateral teleoperation, the slave and the master robots are coupled via a communication network, while the position and force signals are transmitted.
Several continuous-time control schemes have been reported to provide the asymptotic stability of teleoperators despite the constant or variable time delays. The communication delay can destabilize and deteriorate the stability of the teleoperation system.
Besides stability, that is the fundamental requirement for every control system, the teleoperation system must be completely "transparent" [2]. It means that the human operator should feel as if he/she could able to manipulate the remote environment directly. Roughly speaking; transparency can be interpreted as the accurate rendering of the environment to the operator.
The passivity is another important factor of teleoperation system. This attention is due to the well-known property of the passive systems, states that a feedback interconnection of any passive systems is necessarily passive and stable [3]. Passivity is a very useful tool to stabilize a teleoperation system. Its relation to the bilateral teleoperation is that, if the closed-loop teleoperator (master+ slave + communication network) is rendered to be passive, its interconnection with any passive environment and operator will be stable; indeed, considering each part of a teleoperator as a passive system and interconnecting them in a power preserving way leads to having an intrinsically passive system. Also, passivity provides safe interaction and reduces the possible damage on the slave environment and human operator. Nuno *et al.* [4] have studied the passivity of the continuous-time teleoperators with more details.

Several teleoperation control architectures have been proposed in the literature [5], including position-error-based (PEB), direct force reflection (DFR) and 4-channel structure [6]. These architectures are categorized based on the type of signals transmitted between the master and the slave. In theory, four-channel architecture is the most comprehensive structure, because it can be extended to the other architectures through appropriate selection of its control parameters. Furthermore, Yokokohji and Yoshikawa [7] declared that it provides better transparency in compare to the other structures. However, in a recent work, Yang *et al*. [8] have claimed that using finite time control can provide better control performances for teleoperation systems with position error constraints.

Obviously, many studies have concerned the continuous-time control of teleoperation systems. However, these results can neither handle the communication channel discrete nature, nor some of its practical issues (e.g. packet-loss, data-swap). Although bilateral control of teleoperation systems has been discussed a lot, for discrete-time teleoperation systems, much fewer results are available in the literature. This paper is devoted to the sampled-data controlled teleoperation systems to design guidelines to have better transparency and stable teleoperation systems in the sense of motion/force tracking.

Two main approaches can be identified in the literature to design the digital controllers for aforementioned systems [9]: 1-The continuous-time design (CTD): Design is based on the continuous-time model of the system and controller discretization. 2-The discrete-time design (DTD): design is based on the discrete-time model of the system. The CTD is well-established due to the wide range of continuous-time methods that are available for the design of digital controllers. The main drawback of CTD is whether the desired properties of the continuous-time closed-loop system will be preserved, and if so, in what sense for the closed-loop discrete-time controlled system obtained by emulation.

Passivity and stability of sampled data teleoperation systems have not been well-studied such as the continuous ones. This paper presents a review on discrete teleoperation systems, including issues such as stability, passivity and time delays. One of the most important challenges in this area is that the passivity of a discrete teleoperation system is not guaranteed due to energy leaks, which caused by the ZOH (Zero Order Hold). Tavakoli *et al.* [10] investigated a ZOH energy-instilling effect. They represented that the passivity of a teleoperation system can be jeopardized if the continuous-time controllers are substituted with discrete equivalents. Also, Colgate and Schenkel [11] and Gil *et al.* [12] obtained the passivity and stability conditions for the discrete form of a virtual wall. The definition of the passivity for discrete teleoperation systems may be different from that for continuous-time systems. Yoshida *et al.* [13] presented one of the most primary papers that takes into account the discrete-time controllers in teleoperation systems. Both of the time delay and ZOH effects are considered to make the system stable. Using the results of the stability analysis, the robust control structure is proposed for teleoperation systems. Also, Leung and Francis [14] for the first time declared that the passivity property will not be preserved if controllers implement digitally. Loosely speaking, passivity in a discrete-time controlled teleoperation system can be lost or maintained in the face of packet loss and time delays. This depends on the controller mechanism to handle the missing packets. Thus, the stability of the digitally implemented teleoperation systems needs further arguments. This paper is devoted to the sampled-data controlled teleoperation systems and notifies recent contributions in this area.

The rest of this paper is organized as follows: A general comparison between digital and analog control of bilateral teleoperation systems is presented in section 2. In section 3, the preliminaries of the teleoperation systems, including the dynamics of the master and slave robots and hybrid matrix form representation are established. A definition of the transparency which is needed in our review is presented in this section. From the technical point of view, it is better to divide discrete-time controlled teleoperation systems into two categories. The first category relates to works that have been studied the passivity issue (Section 4) and the second category indicates researches concerning the stability of the discrete-time teleoperation systems (Section 5). Also, the effects of discretization methods and network limitations on the stability and transparency are proposed in section 6, and finally the conclusion and future works are given in section 7.

## 2. Discrete-time Control versus Analog Control for Teleoperation Systems

The primary question can be about the necessity of using discrete-time controllers for teleoperation systems. Discrete-time controllers versus continuous-time controllers for teleoperation systems have been discussed in this section. Yang *et al.*[15] presented a comprehensive comparison between digital and analog control of bilateral teleoperation systems in theory and experiments. The constraints of analog controllers for teleoperation systems are highlighted, and guidelines are proposed to address them. Although using continuous-time controllers can improve the transparency of the system compared to its discrete counterpart, analog control implementation difficulties convince us to use discrete equivalents. These troubles include “constraints in the control loop design”, “constraints in the position controller design for a single robot” and “constraints in the haptic teleoperation controller design for two robots”. On the other hand, it is indicated that, for the discrete systems, the large sampling periods, necessitating low control gains to preserve stability and this leads to a low transparency and undesirable task performance.
The mentioned constraints have been applied to the experimental setups and the performance of the discrete-time teleoperation in free motion has been compared to the continuous-time. It is concluded that larger control gains always correspond to the smaller position tracking error for both of them. To have a fair comparison, the human operator's effect can be eliminated using a pulley and rope mechanism. Even in this case, the position tracking errors of the master and slave for continuous-time controllers are less than equivalent discrete-time. Along this paper, Yang *et al.* [16] compared the hybrid parameters of continuous-time and discrete-time controlled bilateral teleoperation systems and the same results of [15] are obtained for performance and stability. In more details, when a discrete-time controller is used for the teleoperation system, the product of sampling time and control gain should be upper bounded to preserve the stability. On the other hand, according to the practical restrictions, the value of the sampling period should be lower bounded. Also, reducing the sampling period can significantly affect the quality of the velocity signal obtained from encoders. This leads to an upper bound on the control gains for the stability maintenance. The stability-imposed upper bound on the gains of the controller degrades the transparency of the teleoperation system. Thus, the human operator cannot complete desirable tasks successfully. Besides all the benefits of analog controllers, they cannot tackle the problem of a discrete-time communication channel with unreliability (e.g., varying-delay, packet-loss and data duplication or swapping).
For another example, consider a bilateral teleoperation system that the slave-environment interaction force signal is measured by a force sensor and after sampling feedback to the operator by a discrete-time controller. When the slave robot penetrates the environment, the transmitted sampled force signal will be less than the real force during each sampling intervals. By contrast, in the free motion, the reflected force signal will be too high compared to reality. Indeed, when the operator performs to probe the passive environment by using the master robot, the energy-instilling discrete-time controller presents the environment to the operator as one emitting energy that causing vibrations. Thus, when we study the passivity and stability of a teleoperator, a discrete-time analysis is necessary to fully account for the CT and DT nature of various signals.
In the next sections, different proposed methods for the discrete teleoperation systems will be introduced.

## 3. System Modelling and Preliminaries

In a bilateral teleoperation system, the slave is operated by an operator through the master and the master receives the force occurring on the slave side. The block diagram of this system is shown in Fig. 2, which is adapted from [17]. The master, the slave, and communication channel are lumped into an LTI (linear time invariant) master-slave two-port network block.

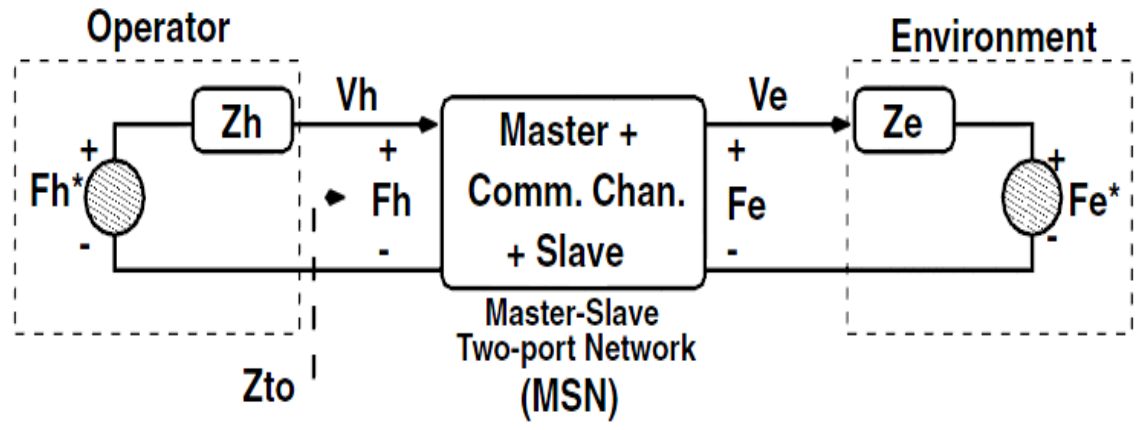

Fig. 2. A bilateral teleoperation network block diagram adapted from [17]

$F_h$ , $F_e$,$V_h$ and $V_e$ are the forces exerted by the operator's hand on the master and by the environment on the slave, and the master and slave velocities, respectively.
Assuming the absence of friction, gravitational forces and other disturbances, the dynamics of the master and the slave robots are given as follows:

$$\begin{aligned} m_m\ddot{x}_m + b_m\dot{x}_m &= F_h - F_m \\ m_s\ddot{x}_s + b_s\dot{x}_s &= F_e - F_s \end{aligned} \tag{1}$$

The subscripts *m* and *s* are used for the master and slave robots, respectively. Likewise, *m* and *b* indicate the mass and the corresponding damping of master and slave robots. The environment and the operator are assumed to be passive and modeled with LTI impedances $Z_e(s)$ and $Z_h(s)$. Control inputs in master and slave sides are represented with $F_m$ and $F_s$. According to the basic sketch of impedance controller that is presented in Fig. 2, the continues-time models of the operator and the environment are:

$$\begin{aligned} F_h &= F_h^* - Z_h(s)sX_m \\ F_e &= F_e^* - Z_e(s)sX_s \end{aligned} \tag{2}$$

In these relations, "$s$" is the Laplace operator and $X_m$ and $X_s$ denote the position of master and slave robots. Also, $F_h^*$ and $F_e^*$ are used for the exogenous force inputs generated by the operator and the environment, respectively. The impedance of the master and slave robots can be modelled by the following equations:

$$\begin{aligned} Z_m &= \frac{1}{m_m s + b_m} \\ Z_s &= \frac{1}{m_s s + b_s} \end{aligned} \tag{3}$$

A hybrid matrix form is used in various papers to model the position error based (PEB) teleoperators. In PEB control, the master robot follows the slave robot position and vice versa. It can be modeled in the hybrid matrix form as:

$$\begin{bmatrix} F_h(s) \\ -sX_s(s) \end{bmatrix} = H(s)\begin{bmatrix} sX_m(s) \\ F_e(s) \end{bmatrix} \tag{4}$$

$$H(s)=\begin{bmatrix} Z_m+C_m\dfrac{Z_s}{Z_s+C_s} & \dfrac{C_m}{Z_s+C_s} \\ -\dfrac{C_s}{Z_s+C_s} & \dfrac{1}{Z_s+C_s} \end{bmatrix} \tag{5}$$

$C_m$ and $C_s$ are controllers used for the master and slave robots, respectively.

For an ideally transparent teleoperation system, regardless of the environment and operator dynamics, we have: $F_h=F_e$ , $\dot{X}_m=\dot{X}_s$.

Thus, from (5) the ideal hybrid matrix can be expressed as $H_{ideal}=\begin{bmatrix}0 & 1\\ -1 & 0\end{bmatrix}$.

The mentioned condition is satisfied if the gains of the controllers are large enough. However, as it will be shown in rest of the paper, this will pose a problem for the stability.

Network theory that has been used for studying electric circuits can be used to model the components of teleoperation systems as depicted in Fig. 2. This figure presents the block diagram of a typical teleoperation system where its components are modeled with a LTI master-slave two-port network (MSN) block.

In the case of a two-port network, the scattering matrix in the frequency domain can be defined based on input $[a_1\ a_2]^T$ and output wave variables $[b_1\ b_2]^T$. The wave variables are defined as

$$a_1=\frac{F_h+R_0V_h}{2\sqrt{R_0}}, a_2=\frac{F_e-R_0V_e}{2\sqrt{R_0}}$$
$$b_1=\frac{F_h-R_0V_h}{2\sqrt{R_0}}, b_2=\frac{F_e+R_0V_e}{2\sqrt{R_0}} \tag{6}$$

$R_0$ is the characteristic resistive impedance of the transmission line[1]. According to the scattering theorem, the relation between the input and output wave variables is defined by the scattering matrix:

$$\begin{bmatrix}b_1\\ b_2\end{bmatrix}=\begin{bmatrix}S_{11} & S_{12}\\ S_{21} & S_{22}\end{bmatrix}\begin{bmatrix}a_1\\ a_2\end{bmatrix}=S(s)\begin{bmatrix}a_1\\ a_2\end{bmatrix} \tag{7}$$

The scattering matrix can be stated in terms of the hybrid matrix:

$$S(s)=\begin{pmatrix}1 & 0\\ 0 & -1\end{pmatrix}(H(s)-I)(H(s)+I)^{-1} \tag{8}$$

Niemeyer and Slotine [18], explored the scattering matrix and its application in teleoperation systems regarding to the energy balance.

## 4. Passivity Analysis of the Discretely Controlled Teleoperation Systems

This section indicates researches that mainly focus on finding conditions to preserve passivity which is one of the important indices to evaluate the performance of the teleoperation systems.

A similar definition can be presented for the passivity of discrete systems like the continuous-time ones. A sampled-data system is said to be passive when its power $P(k)$ consumed up to time $nT$ satisfies the following equation:

$$\sum_{k=0}^{n}P(k)=\sum_{k=0}^{n}x(k)^T y(k)\geq-\gamma \tag{9}$$

If the maximum singular value of scattering matrix is lower than or equal to 1, the system will be passive. The passivity of a LTI system is equivalent to have the system Nyquist diagram entirely in the right half plane.

### 4.1 Passive geometric telemanipulation[19]

Stramigioli [19] presented the interconnection of a discrete-time system with the continuous-time one in a novel approach. This method is based on the port-hamiltonian formalism [20] to describe various parts of the system. Each lumped physical system can be modelled by a port-Hamiltonian system. It is declared that each port-Hamiltonian will be a passive system if the Hamiltonian is bounded from below. A continuous port-Hamiltonian system is discretized while preserving the passivity. Problems such as variable time delays and packet loss have been taken into account in this method. Also, Secchi *et al.* [21] considered power scaling implementation in a discrete port-Hamiltonian based teleoperation system over a packet switched networks. Scaling velocities and forces are in relation with the power which is transmitted between the master and slave sides.

### 4.2 Passivity of the Discrete-time Teleoperation System with Position Error Based Architecture

Passivity condition for the discrete form of the virtual wall is obtained in[11]. Passivity inequality is $b>KT/2+B$, where $T$ is the sampling time, $K$ and $B$ are the stiffness and damping of the virtual wall and $b$ is the haptic interface damping.

Jazayeri and Tavakoli [22, 23] derived sufficient conditions for the passivity of the discrete teleoperation systems with position error-based architectures. This analysis imposes bounds on the damping of slave and master robots, the sampling time and the controller gains. The proposed structure is depicted in Fig. 3. The delay is neglected in this paper. $\tilde{f}_h$ and $\tilde{f}_e$ are exogenous input forces from the operator and the environment, respectively. Two ideal samplers and two ZOHs are used in this structure. $\alpha$ is a position scaling factor.

In [22], the following formula is used to represent the mathematical form of the sampled signal[24]:

$$X^*(s)=L\{x^*(t)\}=\sum_{k=0}^{\infty}x(kT)e^{-skT} \tag{10}$$

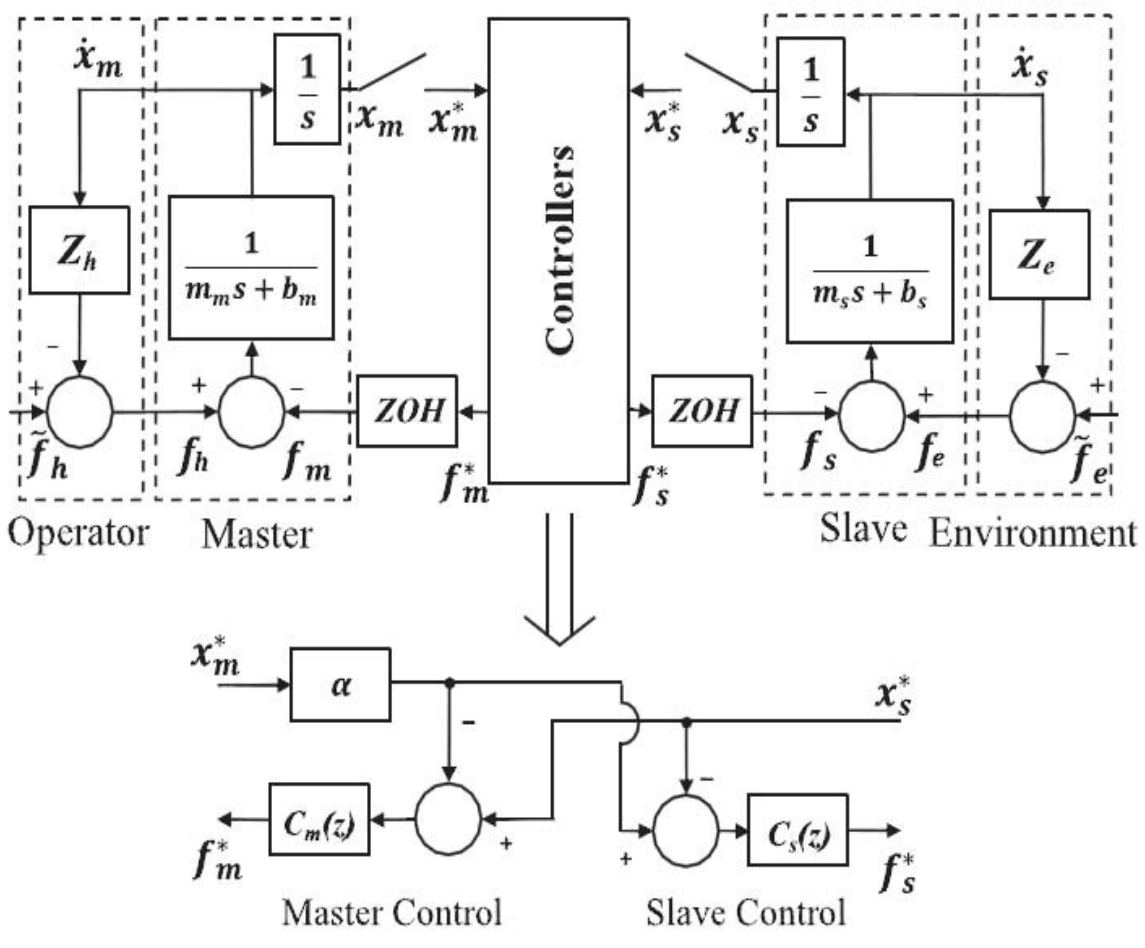

Fig. 3. Structure of teleoperation system with discrete-time controllers and position scaling factor adapted from [22]

Taking into account the ZOH block transfer function, the control inputs can be rewritten as:

$$F_m(s) = \frac{1-e^{-sT}}{sT} F_m^*(s)$$
$$F_s(s) = \frac{1-e^{-sT}}{sT} F_s^*(s) \tag{11}$$

in which, the subscript * means sampled signals.

Defining system error as $e = x_m - x_s$, $\alpha = 1$ and according to the structure of Fig. 3, the discrete-time controllers can be obtained as following laws:

$$F_m^*(s) = C_m(z)\big|_{z=e^{sT}} [X_s^*(s) - X_m^*(s)]$$
$$F_s^*(s) = C_s(z)\big|_{z=e^{sT}} [X_m^*(s) - X_s^*(s)] \tag{12}$$

Using the Parseval's theorem [25], the Fourier transforms and four important lemmas that are stated in [22], sufficient conditions for passivity of a teleoperation system with discrete-time controllers(assuming $\alpha = 1$) are stated as follows:

$$b_m > \frac{T}{1-\cos\omega T} \operatorname{Re}\{(1-e^{-j\omega T})C_m(e^{j\omega T})\}$$
$$b_s > \frac{T}{1-\cos\omega T} \operatorname{Re}\{(1-e^{-j\omega T})C_s(e^{j\omega T})\} \tag{13}$$

For the more general case, $\alpha$ is defined as a position scaling factor and the relation between controllers is stated as

$$C_m(j\omega)/\alpha = C_s(j\omega) = C(j\omega) \tag{14}$$

In this case, the passivity condition is obtained as

$$b_m > \frac{T(\alpha+1)}{2(1-\cos\omega T)} \operatorname{Re}\{(1-e^{-j\omega T})C_m(e^{j\omega T})\}$$
$$b_s > \frac{T(\alpha+1)}{2(1-\cos\omega T)} \operatorname{Re}\{(1-e^{-j\omega T})C_s(e^{j\omega T})\} \tag{15}$$

Mentioned conditions are implemented on a simple PD controller with the following transfer function:

$$C_s(z) = C_m(z) = K + B\frac{z-1}{Tz} \tag{16}$$

Thus, the passivity condition of (13) simplifies to

$$b > KT - 2B\cos\omega T \tag{17}$$

In which $b$ is the minimum of $b_m$ and $b_s$

This condition is held for both of the master and slave robots.

The most noticeable result is that increasing controller gains and sampling time jeopardizes the passivity condition. On the other hand, the physical characteristic of the robot cannot be changed easily. Besides passivity, transparency is another important factor in teleoperation systems. As mentioned in section 2, increasing the controller gains leads to the more transparent system in the continuous-time teleoperation systems. But, in the discrete-time structures, there should be a trade-off between the passivity condition of (17) and transparency of the system. In [22], valuable experimental results using a pair of Phantom Premiums 1.5A robots with JR3 force sensors have been presented to verify the passivity conditions. In [23], along with the [22], the passivity observer is designed in the absence of force sensor, which can be used to detect non-passivity.

## 5. Stability Analysis of the Discretely Controlled Teleoperation Systems

This section provides an overview for the stability of the sampled-data teleoperation systems.

### 5.1 Initial Methods and Experiments

Sheng and Liu [26] categorized the initial studies about the stability of discrete teleoperation systems. Since these methods are a little old, we just will have a brief review on their contents without concerning the formulas and proofs.

Anderson and Spong [27] declared that the stability of the continuous-time bilateral teleoperation system in Fig. 4. is guaranteed for any constant time delay.

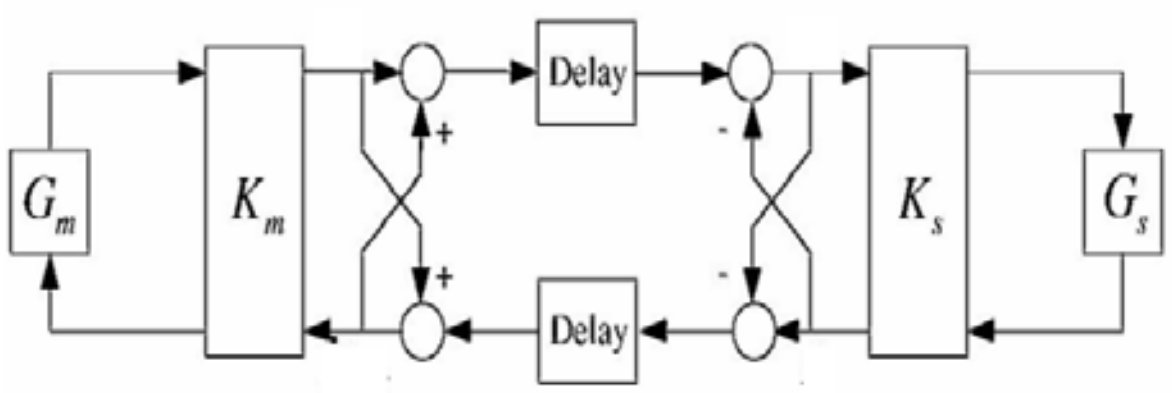

Fig. 4. Continuous-time bilateral teleoperation system adapted from [27]

In this model, $K_m$ and $K_s$ are the controllers of the master side and the slave side, respectively. Also, $G_m$ and $G_s$ are the dynamics of the master robot and slave robot, respectively.

Leung and Francis [28] obtained the sampled-data equivalent of the above teleoperation system. Fig. 5

shows the proposed sampled data structure for the slave side.

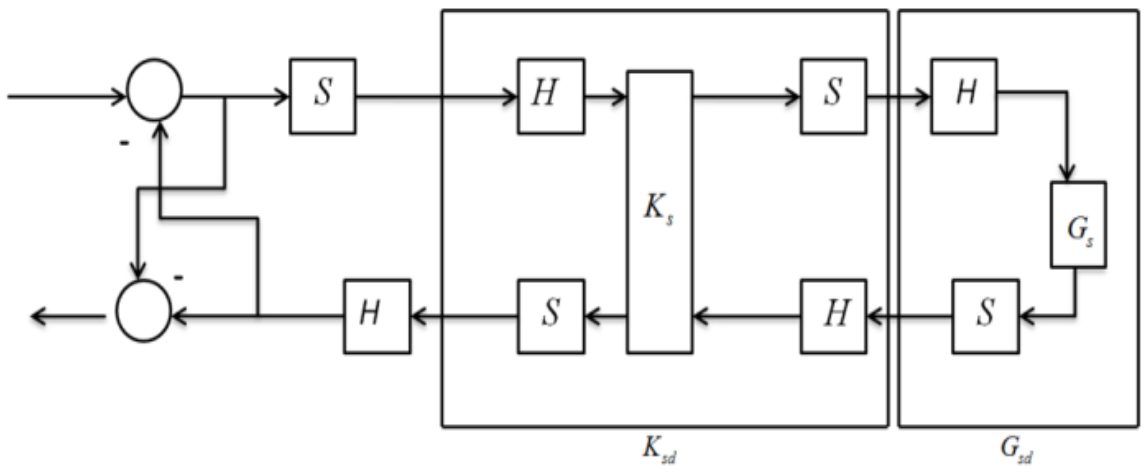

Fig. 5. Proposed discrete bilateral teleoperation system(slave side) adapted from [28]

$G_{sd}$ and $K_{sd}$ are the discrete counterparts of the slave dynamics and its controller, respectively. Also, $S$ and $H$ are used to represent sampler and zero order hold, respectively.

Although passivity has been defined properly both in continuous-time and discrete-time; it is well known that the passivity is lost under discretization. It means that it is possible the discrete-time counterpart of a passive continuous-time system is non-passive. However, we can obtain discrete-time models which preserve the passivity structure. Thus, the proposed discrete form of bilateral teleoperation system in Fig. 5 cannot be passive in the presence of time delay in the communication channel. To tackle this problem, another structure is suggested that is depicted in Fig. 6. In [28], it is claimed that the stability of the discrete teleoperation system in Fig. 5 can be preserved by using six stable, LTI, strictly causal filters in the master and slave sides.

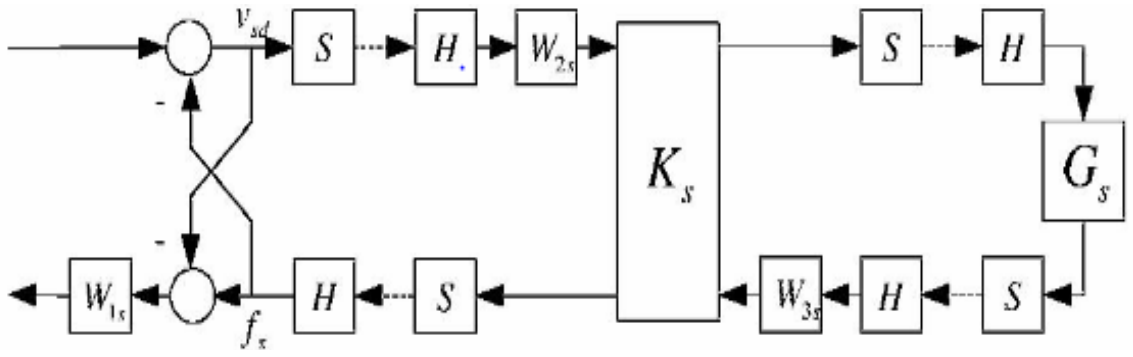

Fig. 6. Discrete bilateral teleoperation systems using low pass filters (slave side) adapted from [28]

$W_{is}\ (i = 1, 2, 3)$ are low path filters at the slave side.

Anderson [29] introduced algebraic instability as a concept to address the problem of discrete-time controller design for teleoperation systems. According to this phenomenon, the sampled-data system has a different passivity and stability behaviour compared to the continuous-time one, which is independent of the sampling time value. Indeed, it is indicated that the common belief that 'sampling fast enough' will cause a discrete system to have the same behaviour of the continuous-time system is invalid for modular teleoperation systems. In the proposed method, the teleoperation system is broken into a series of n-port modules. It is investigated that how each of these sample modules is independently discretized, and how these discrete modules can be combined to achieve the desired teleoperator behaviour. The main point in Anderson's proposed method is using the scattering theory which has been developed to handle problems in the transmission line analysis. Using this theorem, common variables in teleoperation systems such as position signals are transformed into wave variables and the algebraic instability can be overcome. In [29], Tustin's method is used to discretize the continuous scattering operators.

Although this method can be supposed as a first step to solve passivity problem of discrete teleoperation systems, the delay between the slave and the master is assumed to be constant. However, due to the Internet-based teleoperation systems, variable time delays need to be taken into account. Kosuge and Murayama [30] assumed that the communication time delay from the master to the slave and vice versa is different. It is proved that by applying Tustin's method and using scattering theory, the passivity of the whole discrete system can be preserved. In a special case, the sampling rate of communication block assumed to be $N$ times slower than master and slave blocks. This causes degradation in the amount of data, which are transmitted between the master and slave respect to the bandwidth issues in the computer networks. This structure is depicted in Fig. 7.

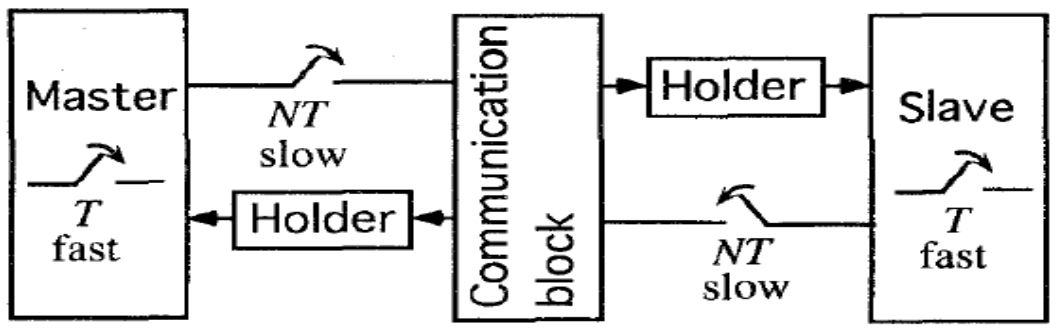

Fig. 7. Discrete-time controlled teleoperation system with different sampling times in communication channel and robots adapted from [30]

Without giving a systematic proof, [30] states that the discrete teleoperation system in Fig. 7, can preserve the passivity under variable time delays and different sampling periods for communication channel and master and slave robots.

Although authors in [30] can cope with the variable time delays, there was no discussion about packet losses. Secchi *et al.* [31] and Chopra *et al.* [32] are the primary researchers which took into account the effect of the packet losses. In the former one, the discrete passivity of the communication channel is considered and concluded that the passivity of the communication channel can be preserved in the presence of packet losses and variable transmission delays. It is declared that in both of the constant and variable time delays the discretized communication channel is lossless and the main difference is that, in constant time delay energy is neither produced or dissipated but simply stored, while for the variable time delay the energy is first dissipated and then injected back to the system. However, in the latter one, it is demonstrated that increasing network delays causes loss of passivity. This depends on the structure of the controller used to cope with missing packets. [33, 34] lead to reconcile the results which are obtained in [31] and [32]. It is shown that passivity in a teleoperation system can be lost or maintained in the presence of packet loss and variable time delays. Indeed, the well-known continuous-time scattering formalism is extended to the discrete domain. It is declared that the passivity maintenance in the presence of variable time delay and packet loss depends on the mechanism used to handle the transmitting packets. Besides the valuable results of [34], the mentioned approach suffered from performance degradation, including loss of force feedback and position drift. Mastellone *et al.* [35] considered the issue and designed a controller that could achieve synchronization and stability. A binary variable $\theta_k \in \{0,1\}$ is utilized to model the information loss in each channel. It means that $\theta_k = 0$ indicates the loss of the packet containing the signal information. The main novelty of the mentioned paper is defining the models of the master and slave similar to the concept of model-based networked control. Comparing with other schemes, this method can overcome the problem of performance degradation. The proposed structure not only achieves passivity but also improves synchronization performance.

### 5.2 Nonlinear method with input-state stability[36]

Consider the following equations that are well-known in nonlinear teleoperation systems:

$$\begin{aligned} M_m(q_m)\ddot{q}_m + C_m(q_m,\dot{q}_m)\dot{q}_m + G_m(q_m) = \tau_m + J_m^T(q_m)\, f_h \\ M_s(q_s)\ddot{q}_s + C_s(q_s,\dot{q}_s)\dot{q}_s + G_s(q_s) = \tau_s - J_s^T(q_s)\, f_e \end{aligned} \qquad (18)$$

In (18), $J$ is the Jacobian matrix. Polushin and Marquez [36] added $\bar{f}_e$ to the right side of the slave equation that represents the delayed force transmitted from the slave to the master with constant delay $\delta_2$. In the other words $\bar{f}_e(t) = f_e(t-\delta_2)$.

Proposed control laws in [36] are as bellows:

$$\begin{aligned} \tau_m = -M_m(q_m)\Lambda_m\dot{q}_m - C_m(q_m,\dot{q}_m)\Lambda_m q_m + \\ G_m(q_m) - K_m(\dot{q}_m + \Lambda_m q_m) \end{aligned} \qquad (19)$$

$$\begin{aligned} \tau_s = -M_s(q_s)\Lambda_s(\dot{\hat{q}}_m - \dot{q}_s) + C_s(q_s,\dot{q}_s)\Lambda_s(\hat{q}_m - q_s) + \\ G_s(q_s) - K_s(\dot{q}_s + \Lambda_s(q_s - \hat{q}_m) \end{aligned} \qquad (20)$$

In (19), $\Lambda_m, \Lambda_s, K_m, K_s$ are arbitrary symmetric positive definite matrices and $\hat{q}_m(t) = q_m(t-\delta_1)$ $\dot{\hat{q}}_m(t) = \dot{q}_m(t-\delta_1)$. $\delta_1$ is the transmission delay from the master to the slave.

**Theorem 1[36]:** For any communication delays $\delta_1, \delta_2 \geq 0$ and considering $(f_h^T, f_e^T)^T$ as inputs of the system (18), using control laws (19) and (20) leads to the input-state stability of teleoperation system.

The proposed controller is discretized based on the following assumptions:

a) All the ZOHs and samplers are synchronized.

b) Both delays from the slave to the master and master to slave are an integer multiple of the sampling time. It means that $\delta_1 = p_1 T, \delta_2 = p_2 T$. $p_1, p_2$ are non-negative integers. Using these assumptions and the structure that is depicted in Fig. 8, it is proved that in the presence of communication delays $\delta_1, \delta_2 \geq 0$, the discretely controlled system will be semiglobally input-state stable.

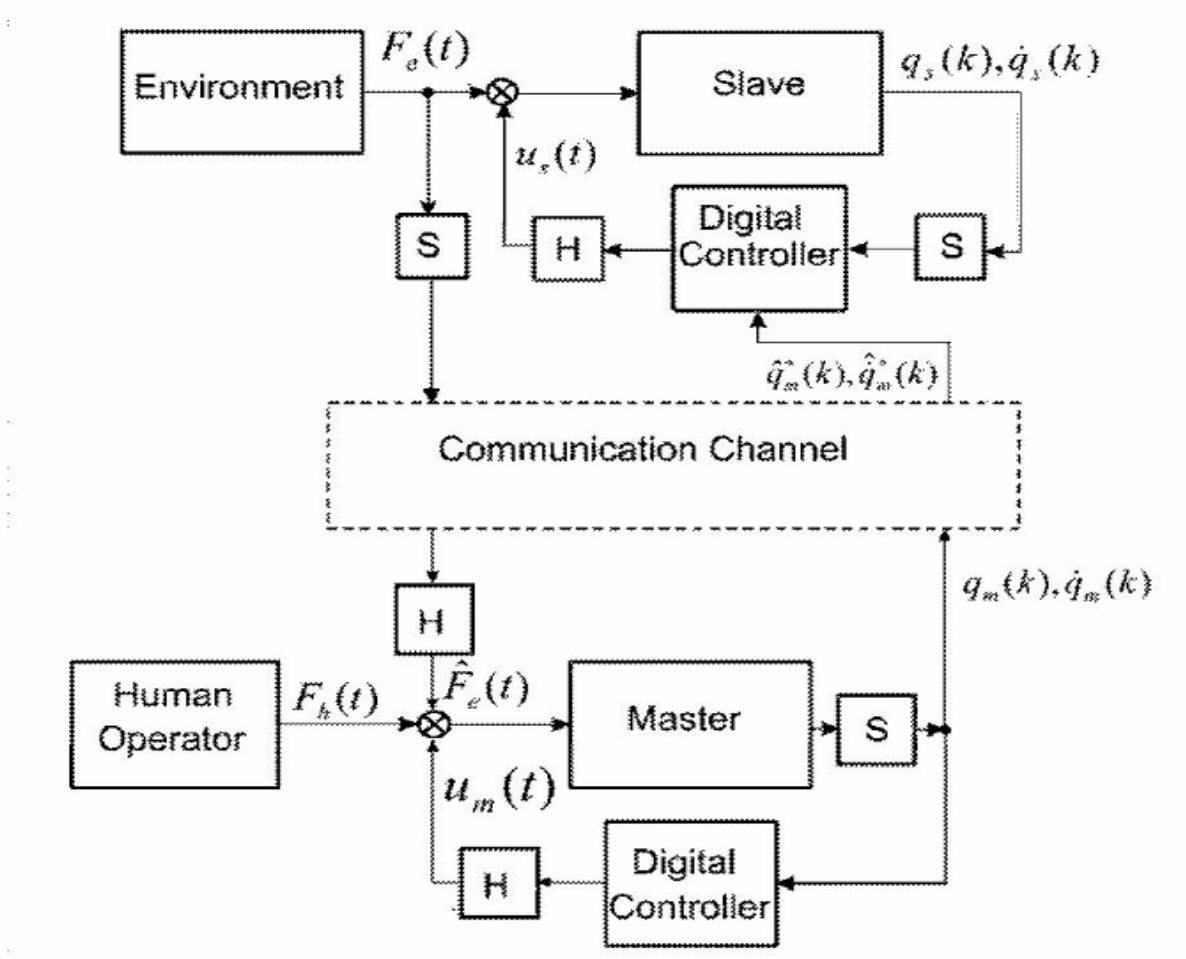

Fig. 8. Bilateral discrete teleoperator structure adapted from [36]

While all the over mentioned conventional bilateral teleoperation researches involve only one pair of local/remote robots, cooperative telerobotics systems can consist of multiple pairs of master/slave manipulators.

A multi-model control strategy is required to enable multiple operators cooperatively manipulate a rigid tool and interact with the task environment via multiple slave robots. Setoodeh *et al.* [37] explored The problem of the cooperative teleoperation control using discrete-time state-space models. Mode-based linear quadratic Gaussian(LQG) controllers are used to deliver stable and transparent responses from each phase of operation. The Nyquist technique has been utilized to analyze the robustness of the controller with respect to the parameters variations.

### 5.3 Stability of the Discrete-time Teleoperation System with Position Error Based Architecture

As a primary work, Hwang *et al.* [38] investigated the sampling rates and inherent time delay effects on the stability of the bilateral teleoperation system that has scaling in position and force signals. This analysis was based on the dynamics of the master and slave robots and the pole location criteria. The most important result of this paper is that increasing one sampling time in both the master and slave robots leads to second-order terms in the dynamics of the system. However, in this work, the communication time delay between master and slave robots is not taken into account because of the short distance.

Absolute stability condition for the discrete form of the virtual wall is obtained in [12]. The stability inequality for the virtual wall is $b > KT/2 - B$. It is obvious that stability condition is less conservative than the passivity one. Also, the effects of nonidealities such as friction, quantization, and energy leaks are taken into account for the stability of the virtual wall in [12, 39, 40].

As mentioned before, using discrete-time controllers may jeopardize the stability of bilateral teleoperation systems. Jazayeri and Tavakoli [41] extended the absolute stability of the sampled-data teleoperation systems using discrete-time controllers. The master, slave, environment and operator assumed to be continuous. Both transmitted position and force signals are subjected by the time delay. This paper imposes bounds for the damping terms of the slave and master, the gains of the controller and the sampling time. Considering bounds for the sampling time is important for the transparency of the system. Larger sampling times require lower control gains to preserve stability. On the other hand, lower control gains can lead to undesirable transparency in the teleoperation systems. A selected architecture for bilateral teleoperation system in [41] is position error based. The exact models of the ZOH and the ideal sampler are taken into account, and the absolute stability of the system is proved using the small gain theorem. In this paper, restrictions of the passivity condition are relinquished. Indeed, in an absolute stable teleoperator, master and slave can be non-passive. Also, operator and environment can have arbitrary passive dynamics and models.

One of the main factors that can jeopardize the stability of the discrete teleoperation system is the energy-instilling behaviour of the ZOH. According to Fig. 3, and equations (10-12), it can be extracted that

$$\begin{aligned} \frac{X_m}{F_m} &= \frac{1}{s}\frac{1}{m_m s + b_m + z_h} \\ \frac{X_s}{F_s} &= \frac{1}{s}\frac{1}{m_s s + b_s + z_e} \end{aligned} \tag{21}$$

Taking into account the transfer function of ZOH, following equations can be found.

$$\begin{aligned} G_m(s) &= \frac{X_m}{F_m^*} = \frac{1}{s}\frac{1}{m_m s + b_m + z_h(s)}\frac{1-e^{-sT}}{sT} \\ G_s(s) &= \frac{X_s}{F_s^*} = \frac{1}{s}\frac{1}{m_s s + b_s + z_e}\frac{1-e^{-sT}}{sT} \end{aligned} \tag{22}$$

Using the model of the sampler (10) and equation (12), the general form of controllers is obtained as

$$\begin{aligned} F_m^*(s) &= C_m(e^{sT})[-\alpha G_m^*(s)F_m^*(s) + G_s^*(s)F_s^*(s)] \\ F_s^*(s) &= C_s(e^{sT})[-G_s^*(s)F_s^*(s) + \alpha G_m^*(s)F_m^*(s)] \end{aligned} \tag{23}$$

In (23), $G_m^*(s)$ and $G_s^*(s)$ are discrete-time forms of $G_m(s)$ and $G_s(s)$.

The characteristic equation of the discrete teleoperation from $\tilde{F}_h$ to any other output is

$$1 + \alpha C_m(e^{sT})G_m^*(s) + C_s(e^{sT})G_s^*(s) \tag{24}$$

In order to analysis of absolute stability, the main theorem is as follows:

**Theorem 2 [41]:** A teleoperation system with discrete-time controllers and position error based architecture is absolutely stable if and only if satisfies the following condition:

$$\left\| M_m N_m + M_s N_s \right\|_\infty < 1 \tag{25}$$

The parameters are defined as:

$$\begin{aligned} N_m &= \frac{\alpha b_s C_m(e^{sT})r(s)}{2b_m b_s + \alpha b_s C_m(e^{sT})r(s) + b_m C_s(e^{sT})r(s)} \\ N_s &= \frac{b_m C_s(e^{sT})r(s)}{2b_m b_s + \alpha b_s C_m(e^{sT})r(s) + b_m C_s(e^{sT})r(s)} \\ M_m &= -1 + \frac{2b_m}{r(s)}G_m^*(s) \\ M_s &= -1 + \frac{2b_s}{r(s)}G_s^*(s) \\ r(j\omega) &= \frac{T}{2}\frac{e^{-j\omega T}-1}{1-\cos\omega T} \end{aligned} \tag{26}$$

Assuming $T_1$ is the delay from the master to the slave and $T_2$ is the delay from slave to master (both of them are integer multiple of sampling time) and considering $\alpha = 0$ the absolute stability condition is stated as follows[41]:

$$\frac{|D + b_s C_m r| + |D + b_m C_s r| + |D|}{\left|2b_m b_s C_m C_s + b_s C_m^2 C_s r + b_m C_s^2 C_m r + D\right|} < 1 \tag{27}$$

In which

$$D = \frac{r^2(1-e^{-(T_1+T_2)s})}{2} \tag{28}$$

For the sake of simplicity, assuming $C_s(z) = n_c C_m(z)$ and $C_m(e^{j\omega T})r(j\omega) = p + jq$ the absolute stability condition for discrete teleoperation systems with PEB architecture can be concluded as

$$\frac{b_m b_s}{\alpha b_s + n_c b_m} > -\mathrm{Re}\{C_m(e^{j\omega T})r(j\omega)\} \tag{29}$$

Comparison of (29) with passivity condition can be instructive. This has been done for a simple PD controller and it is noticed that there is no restricted relation between $\alpha$ as a scaling factor and $n_c$ as the ratio of the master and slave controllers. So, these parameters can be chosen arbitrarily. This point has lots of practical advantages. Even for more simplification and better comparison with the passivity condition, it is assumed that $\alpha = n_c = 1$ and the condition is rewritten as

$$\frac{b_m b_s}{b_m + b_s} > -\mathrm{Re}\{C_m e^{j\omega T})r(j\omega)\} \tag{30}$$

Using a discrete-time PD controller with the following equation:

$$C(z) = k_v(z-1)/T + k_p \tag{31}$$

and assuming that $C_m = C_s = C$ the stability condition is reformed to

$$\frac{b_m b_s}{b_m + b_s} > \frac{k_v T}{2} + k_p \qquad (32)$$

For a given teleoperation system, the left side of the above equation is fixed. This imposes an upper bound for $k_v T$ as the stability condition.

It is obvious that, if $b_m = b_s$, the absolute stability condition will be as same as the passivity condition. But, in the case of $b_m \neq b_s$, the stability condition will be $\frac{1}{b_m} + \frac{1}{b_s} < \frac{-1}{p}$ which is less conservative than the passivity condition.

Some useful remarks are:

*Regarding proposed absolute stability condition, the damping of the master and slave can vary in a less conservative manner.

*This condition allows having an arbitrary position scaling that is useful in the practical setups.

*It is proved that the continuous-time teleoperation system with direct force reflective structure is not absolutely stable if the controller gains are set to be finite. Thus, its sampled-data equivalent cannot be stable too.

The similar analysis can be done for a hybrid model of the discrete-time controlled teleoperation system with 4-channel architecture [10, 42]. Lower and upper bounds on the damping of the controllers and upper bounds on the environment stiffness and the sampling period are obtained to preserve the stability. It is indicated that the sampling period and the maximum admissible stiffness of the environment are in contrast with each other. The proposed discrete architecture is depicted in Fig. 9.

Using Taylor series expansions and Tustin's transformation method, the hybrid model representation of the digitally controlled teleoperation system is obtained.

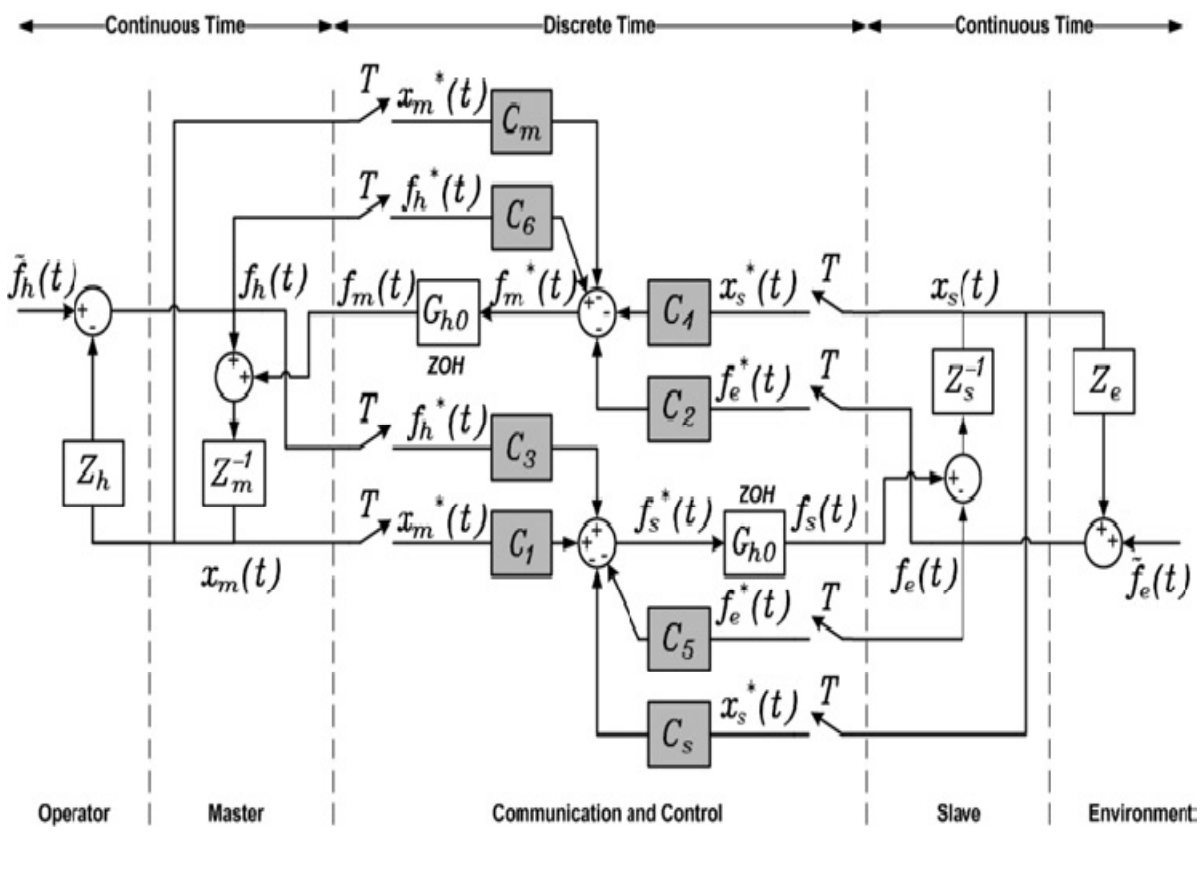

Fig. 9. Discrete-time controlled 4-channel bilateral teleoperation system adapted from [10]

In this structure, $C_s, C_m, C_1, ..., C_6$ are all discrete-time controllers. The stability analysis makes no assumption on $C_2, C_3, C_5, C_6$, in order to cover all teleoperation architectures including DFR and PEB.

### 5.4 Stability Conditions for a Non-Passive Operator or Environment

Recent researches try to get rid of the passivity conditions of the operator or the environment. Miandashti and Tavakoli [43] used discrete-time circle criterion for the absolute stability of a sampled-data haptic interaction. The main advantage of this method is that the operator is allowed to have an active behaviour. The well-known Colgate's stability condition for a teleoperation system with one user is extended to the multi-user case, while allowing all of the operators to have passive or active roles. Assuming the impedance of the environment has been approximated as $K + [\{B(z-1)\}/Tz]$, four different cases for the operator's passivity are considered. These states include:

1- **Passive operator with no delay**. In this case, like pervious result in section 4.1, the stability condition is obtained as:

$$b > \frac{KT}{2} + B \qquad (33)$$

2- **Passive operator with Time delay.** For a delayed teleoperation system, the stability condition will take a different form depending on the environment model. The stability condition will be derived as follows:

$$b + B > \frac{KT}{2} + Kt_d \qquad (34)$$

In which $t_d$ is a constant time delay.

3- **Active operator with no delay.** Assume that the system is active with the shortage of passivity $z_a$. Extending the results of [11] and after replacing $b$ by $b - z_a$ the stability condition will be

$$b - z_a > \frac{KT}{2} + B \qquad (35)$$

This condition allows active intervention for the operators.

4- **Active operator with time delay.** In the case of delayed teleoperation systems in which the operator has a shortage of passivity $z_a$ the stability condition will be

$$b - z_a + B > Kt_d + \frac{KT}{2} \qquad (36)$$

## 6. Study the Effect of Discretization Methods and Network Limitations on the Stability and Transparency

Another important point that should be taken into account is the method of discretization of the position and velocity signals. This can affect the passivity of teleoperation system or haptic interface.

Yin *et al.* [44] and Haddadi *et al.*[45] assumed that both the position and velocity signals are sampled. Consider that $H(z)$ denotes the transfer function of the simulated virtual environment. Using the backward difference (BD) discretization method to digitally implement a damper-spring virtual environment in a position-sampled system, the transfer function will be $H(z)=K+B\frac{1-z^{-1}}{T}$, where $K$ and $B$ are the stiffness and the damping of the environment. For Tustin implementation the transfer function becomes $H(z)=K+\frac{2}{T}B\frac{1-z^{-1}}{1+z^{-1}}$. The block diagram of the position-sampled system is depicted in Fig. 10. $X_h, V_h, X_d, V_d, F_e, F_h$ are the master position and velocity, and the corresponding sampled position and velocity, environment force and operator's force on the master, respectively.

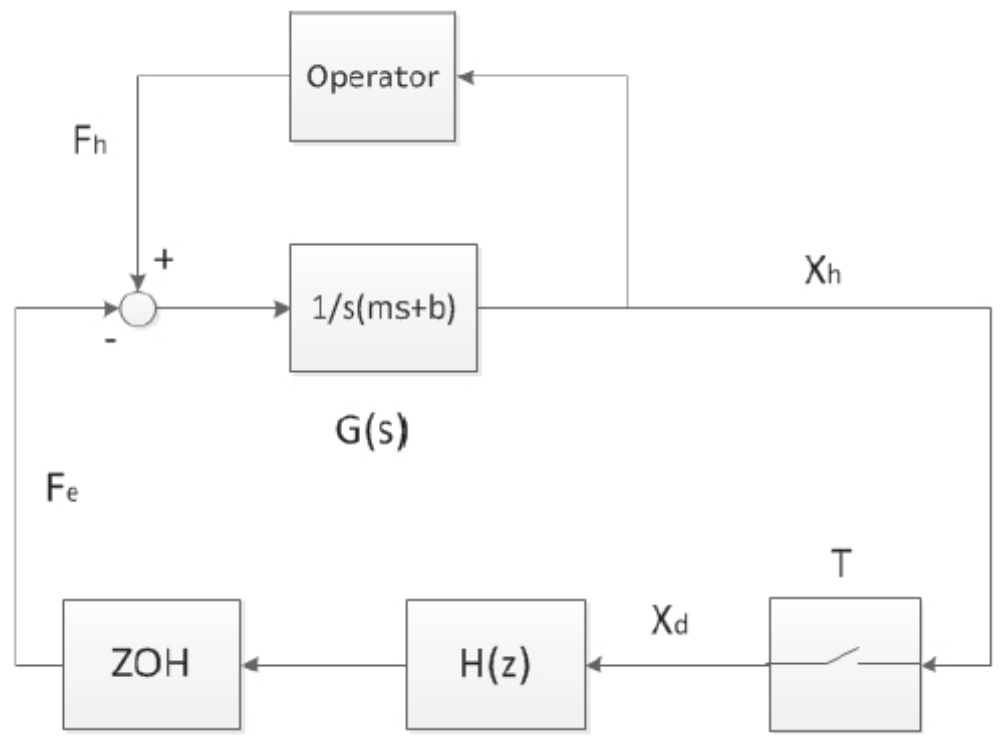

Fig. 10. Block diagram of position-sampled system coupled to an operator adapted from [44]

For the velocity sampled systems, $H(z)=\frac{KT}{1-z^{-1}}+B$ for the BD approach, and $H(z)=\frac{KT}{2}\frac{1+z^{-1}}{1-z^{-1}}+B$ for the Tustin method. The block diagram of velocity-sampled system is depicted in Fig. 11.

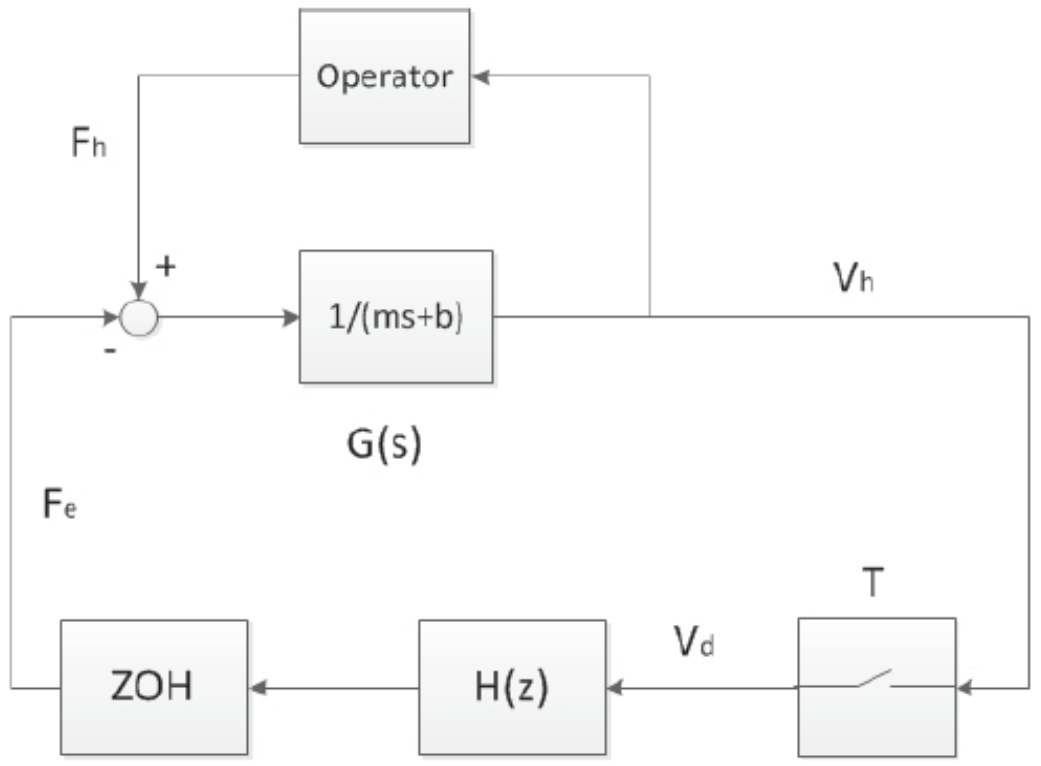

Fig. 11. Block diagram of velocity-sampled system coupled to an operator adapted from [44]

The characteristic equation of the uncoupled system is used as the building block for the stability analysis and the aforementioned methods are considered to implement a linear-time-invariant damper-spring environment. This study can be valuable since the type of the sampled signals and the discretization method can make a considerable effect on the stability of teleoperation systems. The results present that sampling the velocity signals, using backward difference method; will increase the stability range of the environment dynamics. It is also experimentally represented that sampling both of the position and velocity signals, provides a larger stability region.

Although these researches give a valuable comparison between different methods of discretization, they do not take into account the limitations of signal transmission and communication channel. The transmitted signal bandwidth is generally limited to the Nyquist frequency determined by the packet transmission rate. The authors in [46-48] make their efforts to solve these problems.

Mizuochi and Ohnishi [46] proposed a method to transmit high-frequency signals even under severe limitations on the sampling time and packet transmission rate. A low-pass filter with the discrete Fourier transform is used for coding of signals, and inverse discrete Fourier transform is used for the decoding. Although the whole idea seems to be simple, the valuable advantage is achieved by extracting the signals as frequency spectra. Kubo and Ohnishi [47] studied bandwidth compression and transmission of environmental information with more details. In this paper, discrete Fourier transform matrices are used to convert the position and force signals of the slave side into the environmental modes. The main drawback of this research is that only the low-frequency components of spatial information have been considered. However, the high-frequency components are more important for certain tasks.

Huang and Dongjun [48] assumed that the master and slave robots are nonlinear continuous-time systems with sampled-data controllers. The communication channel is the discrete-time packet-switching Internet line, including unreliabilities such as varying time delays and packet-loss. In this paper, a novel virtual-proxy based hybrid teleoperation control framework is introduced to pacify the communication channel unreliabilities. This controller can also guarantee position coordination and force reflection.

***An experimental point:**

Yang *et al.* [49] designed a field programmable analog arrays (FPAA) based controller for a bilateral teleoperation system. FPAA is a modified type of reconfigurable analog circuits. It can achieve higher transparency and better performance for a PEB teleoperation system. But what is the purpose behind this claim? It is experimentally represented that the FPAA-based controller can solve the low control gains issue caused by large sampling periods in digitally controlled teleoperation systems. Because of its inner features, it behaves as the analog control system and fundamentally eliminates the limitation brought by the sampling period.

In another more recent experimental work, Wrock and Nokleby [50] have presented a novel command strategy to combine human operator skill with haptic feedback to achieve better performance than previously attainable using only a single input device.

## 7. Conclusion and Future Works

Although bilateral control of teleoperation systems have been discussed a lot, there are a few materials about discretely controlled teleoperation systems in the control literature. Passivity and stability of the sampled data teleoperation systems have not been well-studied such as the continuous ones. This paper presented a review on the discrete teleoperation systems, including issues such as stability, passivity, and transparency.

Discretization does not preserve passivity and absolute stability of teleoperation systems due to energy leaks caused by ZOH. Absolute stability needs to impose bounds on the inner parameters of the teleoperation system and its controllers. This can provide guidelines to design controllers with high gains without losing stability and transparency.

For future works on the discrete-time controlled bilateral teleoperation systems, passivity and absolute stability conditions should be taken into account and extended for the 4-channel control architectures. In addition, the time delay between the master and slave should be considered. The implementation problem for the case that the sampling rate of the communication block is lower than the rate of the master and slave robots should be taken into account. Mixing the capabilities of the discrete-time and continuous-time controllers to achieve high transparency and stable teleoperation systems involving both soft and hard environments can be considered as another future work. These hybrid systems can benefit from the better transparency of the continuous-time controllers and flexibility of discrete-time ones.

Using sampled-data design (SDD) to represent the inter-sample behaviour of the transmitted signals of the teleoperation system is another important point in the design procedure. Additionally, most of the mentioned methods belong to the CTD control categories, and DTD controllers require more attention.